\documentclass{elsart}
\usepackage{graphics}

\begin{document}
\begin{frontmatter}
\title{One-electron bands, quantum Monte Carlo, and real superconductors}
\author{Erik Koch\thanksref{mail1}}
and
\author{Olle Gunnarsson}
\address{Max-Planck-Institut f\"ur Festk\"orperforschung\\ 
         Heisenbergstra\ss e 1, 70569 Stuttgart, Germany}
\thanks[mail1]{email: koch@and.mpi-stuttgart.mpg.de}

\begin{abstract}
We use the doped Fullerenes as an example of how realistic systems can
be described by simple models. Starting from the band structure we set up
a tight-binding model that describes the $t_{1u}$ conduction band.
Adding correlation terms we arrive at a generalized Hubbard Hamiltonian
that we treat using quantum Monte Carlo. To address the problem of 
superconductivity in the doped Fullerenes, we study the screening of a
point charge. We find surprisingly efficient metallic screening even for 
strong correlations, almost up to the Mott transition, and discuss the
implications on superconductivity, in particular the effect of the 
efficient screening on the Coulomb pseudopotential and the 
electron-phonon coupling.
\end{abstract}


\end{frontmatter}

\section{Introduction}

The sheer size of the many-body Hilbert space makes treating strongly 
correlated systems adequately extremely difficult or even impossible.
Examples of considerable interest are the superconducting doped Fullerenes. 
Even for a single C$_{60}$ molecule a full many-body calculation is still a 
challenge and calculations for solids made of Fullerenes are simply out of
question. In this situation we are forced to restrict our attention to only 
the most relevant degrees of freedom. For the doped Fullerenes these are the 
electrons in the $t_{1u}$-band. Starting from {\it ab initio} density functional
calculations we set up a tight-binding Hamiltonian that describes the 
electrons in the $t_{1u}$-band only. Including correlation effects we arrive 
at a generalized Hubbard Hamiltonian that can be treated by quantum Monte Carlo
(QMC). Using the fixed-node approximation we study the screening of an point 
charge by the electrons in the $t_{1u}$-band. We find that the screening is 
surprisingly efficient even for strong correlations, almost up to the Mott 
transition. This has important implications for superconductivity in the 
doped Fullerenes. Given that molecular vibration energies are of the same order
as electronic energies, retardation effects are inefficient at reducing the 
electron-electron repulsion. It is therefore not clear how the weak 
electron-phonon attraction can lead to superconductivity. Efficient metallic 
screening, as found in our calculations, can, however, reduce the 
electron-electron repulsion enough to allow for an electron-phonon driven 
superconductivity. But the screening does, of course, also affect the 
coupling to the phonons. It turns out that, due to screening, the alkali and 
$A_g$ modes couple only weakly, while the coupling to the $H_g$ modes is
not affected. Therefore, although being electron-phonon driven, 
superconductivity in the doped Fullerenes differs in important ways from the 
conventional picture of superconductivity.

\section{Model Hamiltonian}

Treating correlations in the Fullerenes is quite difficult. Even for a
single C$_{60}$ molecule a full many-body calculation is still a challenge
and simulations of Fullerenes, which are solids made of C$_{60}$ molecules
are well beyond current computational capabilities. Solid C$_{60}$ is
characterized by weak inter-molecular interactions. Hence the molecular 
levels merely broaden into narrow, well separated bands \cite{erwin}. Doping 
the solid with alkali metals has the effect of filling the band originating 
from the molecular $t_{1u}$-level with the weakly bound valence electrons of 
the alkali atoms. Since the $t_{1u}$-level is three-fold degenerate the 
corresponding band can hold up to six electrons per molecule; for A$_3$C$_{60}$,
e.g., the band is half-filled. When we are interested in the low-energy 
properties of the $t_{1u}$ electrons, it is a good approximation to focus only 
on the region around the Fermi level; i.e.\ on the $t_{1u}$-band, projecting 
out all the other bands \cite{lowdin}. That way we arrive at a tight-binding 
Hamiltonian comprising only the $t_{1u}$ orbitals, that reproduces the 
{\it ab initio} band structure remarkably well \cite{nato}.

To obtain a realistic description we have to include the correlation 
arising from the Coulomb repulsion among the electrons on the same molecule. 
The resulting Hamiltonian describes the interplay between the hopping of 
electrons between different molecules and the Coulomb repulsion among the
electrons on the same molecule
\begin{equation}\label{Hamil}
 H=\sum_{\langle ij\rangle} \sum_{nn'\sigma} t_{in,jn'}\;
              c^\dagger_{in\sigma} c^{\phantom{\dagger}}_{jn'\sigma}
 +\;U\sum_i\hspace{-0.5ex} \sum_{(n\sigma)<(n'\sigma')}\hspace{-1ex}
       n_{i n\sigma} n_{i n'\sigma'} .
\end{equation}
Here $c^\dagger_{in\sigma}$ creates an electron of spin $\sigma$ in the
orbital with index $n\in\{1,\ldots,3\}$ on molecule $i$, the $t_{in,jn'}$ are
hopping matrix elements between orbitals $n$ and $n'$ on neighboring 
molecules, and $n_{in\sigma}=c^\dagger_{in\sigma}c_{in\sigma}$ are
occupation operators. Varying the value of the interaction term $U$, we
can study the effect of correlations. The physical value for the doped
Fullerenes is $U\approx1.2-1.4\,$eV \cite{expU}, which has to be compared
to the width of the $t_{1u}$-band, $W\approx0.5-0.85\,$eV.

\section{Quantum Monte Carlo}

We now give a very brief outline of the quantum Monte Carlo method for
determining ground states. The basic idea is to use the Hamiltonian to project
out the ground-state from some trial function $|\Psi_T\rangle$, that we have 
guessed:
\begin{equation}
  e^{-\tau\,H}\,|\Psi_T\rangle \stackrel{\tau\to\infty}{\longrightarrow}
  |\Psi_0\rangle .
\end{equation} 
To see how this works let us assume we knew the expansion of the trial function
in eigenfunctions of $H$
\begin{equation}
  |\Psi_T\rangle=\sum_n c_n|\Psi_n\rangle
  \quad\Rightarrow\quad 
  e^{-\tau\,H}\,|\Psi_T\rangle=\sum c_n\,e^{-\tau E_i}\,|\Psi_n\rangle .
\end{equation}
Thus the component with the lowest eigenenergy is damped least during the 
projection, i.e., if $c_0\ne0$, in the limit of large $\tau$
the ground-state component will dominate. In practice we use for  
Hamiltonians $H$ with a spectrum that is bounded, both, from below {\em and} 
above, an iterative projection of the form \cite{DMC}
\begin{equation}\label{qmciter}
 |\Psi^{(0)}\rangle=|\Psi_T\rangle
 \qquad 
 |\Psi^{(n)}\rangle=[1-\tau(H-w)]|\Psi^{(n-1)}\rangle ,
\end{equation}
which, for small but finite $\tau$, also exactly projects to the ground-state.
We see that the basic operation in (\ref{qmciter}) is a matrix-vector product.
Since we are working in a many-body Hilbert space, the dimension of the 
vectors is, however, in general enormous; see Table \ref{dimH} for an
illustration.

To understand the Monte Carlo method for doing the iteration we first rewrite 
(\ref{qmciter}) in configuration space. Here $R$ denotes a configuration of
the electrons in real space  
\begin{equation}
  \sum |R'\rangle\langle R'|\Psi^{(n)}\rangle
    = \sum_{R,R'} |R'\rangle
       \underbrace{\langle R'|1-\tau(H-E_0)|R\rangle}_{=:F(R',R)}
       \langle R|\Psi^{(n-1)}\rangle .
\end{equation}
We see that the matrix $F(R',R)$ maps configuration $R$ into configurations
$R'$. We clearly cannot follow every possible new configuration $R'$ since 
that would lead to an exponential growth in the number of configurations as 
we iterate. The idea of Monte Carlo is then to sample {\em only one} of the
configurations $R'$ with a probability $p(R',R)$. To do that we want to 
interpret the matrix elements of $F(R',R)$ as probabilities. They are, however,
in general not normalized and can even be negative
\begin{equation}
 F(R',R) =\underbrace{p(R',R)}_{\rm prob.}\;\;
          \underbrace{m(R',R)}_{\rm norm.\& sign} . 
\end{equation}
Normalization introduces the need for population control, while negative
matrix elements introduce the sign problem \cite{FNDMC,appl}.

\begin{table}
\begin{center}
\begin{tabular}{rrll}
   $N_{\rm mol}\hspace*{-0.5ex}$ & \multicolumn{2}{c}{dimension} &
   \multicolumn{1}{c}{\hspace*{-2ex}memory/GB}\\
   \hline\\[-4ex]
     4 & $     853\;776 =\!$ & $8.5\cdot 10^{  5}$ & $3.2\cdot 10^{ -3}$\\[-1ex]
     8 & $7\;312\;459\;672\;336 =\!$
                             & $7.3\cdot 10^{ 12}$ & $2.7\cdot 10^{  4}$\\[-1ex]
    16 &                     & $1.0\cdot 10^{ 27}$ & $3.9\cdot 10^{ 18}$\\[-1ex]
    32 &                     & $4.1\cdot 10^{ 55}$ & $1.5\cdot 10^{ 47}$\\[-1ex]
    64 &                     & $1.3\cdot 10^{113}$ & $4.9\cdot 10^{104}$\\[-1ex]
   108 &                     & $2.3\cdot 10^{192}$ & $8.5\cdot 10^{183}$\\
   \hline
\end{tabular}
\end{center}
\vspace{3ex}
\caption[]{\label{dimH}
           Dimension of the Hilbert space for a system with 
           $N_\uparrow+N_\downarrow$ electrons on $M$ lattice sites.
           For our model of doped Fullerenes $M=3\,N_{\rm mol}$.
           The last column gives the amount of memory needed to store
           a single state vector.}
\end{table}

\section{Screening of a Point Charge}

In conventional superconductors the electron-phonon interaction leads to an
effective electron-electron attraction. This attraction is, of course,
counteracted by the Coulomb repulsion between the electrons. In the conventional
picture this repulsion is, however, strongly reduced by retardation effects
\cite{bcs}. The resulting effective Coulomb repulsion is described by the 
dimensionless Coulomb pseudopotential $\mu^\ast$, which is believed to be of
the order of $0.1$. For the doped Fullerenes the situation is different:
Retardation effects are inefficient \cite{rmp}. Therefore the screening of the 
Coulomb interaction becomes important for reducing the electron-electron 
repulsion. Assuming that the random phase approximation (RPA) is valid for
the electrons within the $t_{1u}$-band it was found that efficient metallic
screening significantly reduces the Coulomb pseudopotential \cite{rpa}.
In this scenario the Coulomb pseudopotential $\mu^\ast\approx 0.3$ is
substantially larger than that for conventional superconductors, but it is not
too large to prevent superconductivity from being driven by the 
electron-phonon interaction. For strongly correlated systems like 
the doped Fullerenes the use of the RPA seems, however, highly questionable.

\begin{figure}
\centerline{\rotatebox{270}{\resizebox{!}{0.7\textwidth}{\includegraphics{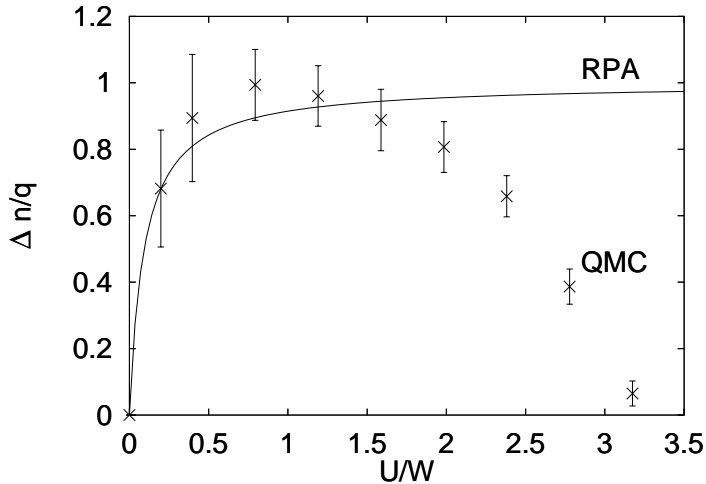}}}}
\vspace{3ex}
\caption[]{\label{scr}
           Screening charge $\Delta n$ on the site of a test charge $q=1/4$ 
           electron charges as a function of $U/W$, extrapolated to infinite
           cluster size.
          }
\end{figure}

To address this question, we have investigated how well the RPA describes the 
screening within the $t_{1u}$-band. It is clear that the RPA properly describes
the screening when the kinetic energy is much larger than the interaction 
energy, i.e.\ RPA works well in the weakly correlated limit. For strong 
correlations, where the Coulomb energy dominates, the RPA gives qualitatively 
wrong results:
Introducing a test charge $q$ on a molecule, RPA predicts that almost the same
amount of electronic charge moves away form that molecule, since for a Coulomb
integral $U$ much larger than the band width $W$ the gain in potential energy 
dominates the cost in kinetic energy. The RPA neglects, however, that in this
limit, when an electron leaves a molecule it has to find another molecule with
a missing electron, or there will be a large increase in Coulomb energy.
It is not clear for what value of the Coulomb interaction $U$ this qualitative 
breakdown of the RPA starts, and up to which values of $U$ RPA gives still
accurate results. To find out we have performed quantum Monte Carlo 
calculations. To study the screening of a point charge $q$ on molecule with
index $c$ we consider the Hamiltonian
\begin{equation}\label{Hamilq}
 H=\sum_{\langle ij\rangle} \sum_{nn'\sigma} t_{in,jn'}\;
              c^\dagger_{in\sigma} c^{\phantom{\dagger}}_{jn'\sigma}
 +\;U\sum_i\hspace{-0.5ex} \sum_{(n\sigma)<(n'\sigma')}\hspace{-1ex}
       n_{i n\sigma} n_{i n'\sigma'}
 +q U\sum_{n\sigma} n_{cn\sigma} ,
\end{equation}
which differs from (\ref{Hamil}) only by the last term describing the 
interaction with the test charge. As a trial function we use a generalized 
Gutzwiller function of the form
\begin{equation}
  |\Psi_T\rangle = g^D\, g_0^{n_c}\;|\Phi_0\rangle ,
\end{equation}
where $|\Phi_0\rangle$ is a Slater determinant, $D$ is the number of 
double occupancies in the system, and $n_c$ is the number of 
electrons on the molecule with the test charge. Calculating the expectation 
value $n_c(VMC)=\langle\Psi_T|n_c|\Psi_T\rangle$ by variational Monte Carlo 
and the mixed estimator $n_c(DMC)=\langle\Psi_T|n_c|\Psi_0\rangle$ by 
fixed-node diffusion Monte Carlo, we obtain the ground-state expectation 
value $n_c=\langle\Psi_0|n_c|\Psi_0\rangle$ from the extrapolated estimator 
$n_c\approx2n_c(DMC)-n_c(VMC)$. To estimate the accuracy of our approach we 
have compared the results of the quantum Monte Carlo (QMC) calculations with the 
results from exact diagonalization for a cluster of 4 molecules. We find that 
the QMC calculations are accurate up to very large values of the Coulomb 
interaction $U$. Performing QMC calculations for clusters of sizes 
$N_{\rm mol}=$ 
32, 48, 64, 72, and 108 molecules, where exact diagonalization is not possible 
(cf.\ Table \ref{dimH}), we have extrapolated the screening charge
$\Delta n_c=n_c(0)-n_c(q)$ to infinite cluster-size, assuming a finite-size
scaling of the form $\Delta n_c=\Delta n_c(N_{\rm mol})+\alpha/N_{\rm mol}$
\cite{screening}. The finite-size extrapolation gives only a small correction 
to the screening charge found for the larger clusters. Results for $q=0.25\,e$ 
are shown in Figure \ref{scr}. For small values of $U$ the RPA somewhat 
underestimates the screening, a behavior also found in the electron gas 
\cite{hedin}. For intermediate values of $U$ ($U/W\approx1.0-2.0$) the RPA 
still gives surprisingly accurate results, while for larger $U$ it rapidly 
becomes qualitatively wrong, as discussed above. We thus find efficient, 
RPA-like screening even for quite strong correlations close to the Mott 
transition.

\section{Screening and Electron-Phonon Coupling}

As pointed out in the preceeding discussion, efficient metallic screening
helps to reduce the effective electron-electron repulsion, i.e.\ the Coulomb
pseudopotential. But the screening also affects the electron-phonon coupling.
At first it might appear that efficient screening is not really helpful for
superconductivity. Phonons couple to the electrons by perturbing the potential
seen by the electrons. An example is the longitudinal mode of a jellium.
Efficient screening tends to weaken the coupling to such phonons, since it
reduces the perturbation considerably: The bare coupling constant $g$ is
reduced to $g/\varepsilon$, where $\varepsilon$ is the dielectric 
constant \cite{bcs}. To some extent, such a reduction is also at work 
in the Fullerenes. An example are the alkali phonons. Each C$_{60}$ molecule 
is surrounded by 14 alkali ions that are bound with quite weak force constants. 
They should therefore respond strongly when an electron arrives on a C$_{60}$ 
molecule. This was, however, not confirmed experimentally; an alkali isotope
effect could, e.g., not be observed \cite{isotope}. Given the efficient metallic
screening this finding can be naturally understood. When an electron arrives 
on a C$_{60}$ molecule, other electrons leave the molecule, which thus stays
almost neutral. The alkali ions then see only a small change in the net charge
and therefore couple weakly.

\begin{figure}
\centerline{\resizebox{0.7\textwidth}{!}{\includegraphics{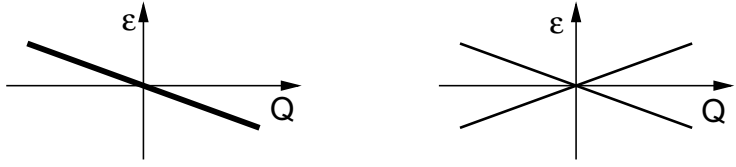}}}
\vspace{3ex}
\caption[]{\label{elphc}
           Schematic picture of the change in the energy levels as function
           of the phonon coordinate $Q$ (deformation of molecule). The left
           picture shows a phonon that shifts the center of gravity of the
           energy levels, while in the right picture the center of gravity
           is conserved.}
\end{figure}

To analyze the situation more closely, we consider the change in the energy
of the molecular orbital $\alpha$ under a deformation of the molecule with 
amplitude $Q$
\begin{equation}
 \Delta\epsilon_\alpha^0 = g_\alpha\,Q ,
\end{equation}
where $g_\alpha$ is the electron-phonon coupling. An illustration is given in
Figure \ref{elphc}. The change of the on-site energy $\epsilon_\alpha^0$ will
induce a response of the $t_{1u}$ electrons. Since the effect of a point
charge $q$ is just to shift the on-site energy by $q\,U$ (cf.\ (\ref{Hamilq})),
the screening charge induced by the change $\Delta\epsilon_n^0$ is given by
\begin{equation}
  \Delta n_\alpha = -\gamma\;{\Delta\epsilon_\alpha^0\over U},
\end{equation}
where $\gamma>0$ measures the efficiency of the screening: 
$\Delta n=-\gamma\,q$. The total screening charge induced by the molecular
deformation $Q$ is then given by
\begin{equation}
  \Delta n = \sum_\alpha \Delta n_\alpha
           = -{\gamma\over U}\,\left(\sum g_\alpha\right)\,Q .
\end{equation}
Including screening, the shift in the molecular levels is then given by
\begin{equation}
  \Delta\epsilon_\alpha = \Delta\epsilon_\alpha^0 + U\sum_\beta \Delta n_\beta 
                        = \left(g_\alpha-\gamma\sum_\beta g_\beta\right)\;Q .
\end{equation}

For molecular solids like the doped Fullerenes the electron-phonon coupling
is given by the dimensionless constant \cite{lambda}
\begin{equation}
  \lambda \propto \sum_\alpha\left({\Delta\epsilon_\alpha\over Q}\right)^2
                = \sum_\alpha\left(g_\alpha-\gamma\sum_\beta g_\beta\right)^2 .
\end{equation}
Given efficient metallic screening $(\gamma\approx1)$, the coupling to 
phonons that cause a net shift of the molecular levels $\sum g_\beta\ne0$
will be reduced, while modes that leave the center of gravity of the
molecular levels unchanged $(\sum g_\beta=0)$ will not be affected. 
Such modes are the $H_g$ modes in C$_{60}$. For these modes efficient 
screening serves to reduce the electron-electron repulsion without affecting
the electron-phonon coupling.

\section{Conclusion}

We have described how to construct a model for the $t_{1u}$ electrons in
the doped Fullerenes, that can be analyzed by many body techniques. Using
quantum Monte Carlo we have calculated the static screening of a point charge.
We find that the RPA works surprisingly well, almost up to the Mott transition.
The metallic screening helps to reduce the electron-electron repulsion in
the doped Fullerenes, where retardation effects are inefficient. But the
screening in general also reduces the electron-phonon coupling. In a molecular
solid there can be, however, intra-molecular modes that are not screened, 
examples being the $H_g$ modes in the Fullerenes. We thus find that, although 
superconductivity in the Fullerenes is driven by the electron-phonon coupling,
it differs in important points from the textbook picture of superconductivity.

\begin{ack}
This work has been supported by the Alexander-von-Humboldt-Stiftung under
the Feodor-Lynen-Program and the Max-Planck-Forschungspreis.
\end{ack}

\newcommand{\jref}[4]{{\it #1} #2 (#4) #3}
\newcommand{\prl}{Phys.\ Rev.\ Lett.\ }
\newcommand{\prb}{Phys.\ Rev.\ B}

\end{document}